%% file: samplepaper.tex
\documentclass[runningheads]{llncs}
\usepackage[T1]{fontenc}
\usepackage{newunicodechar}

\usepackage{tikz}
\usetikzlibrary{arrows.meta}
\usepackage{newtxtext}       %
\usepackage[varvw]{newtxmath}       %
\usepackage{xspace}
\usepackage{booktabs,hyperref}
\usepackage{listings}
\usepackage{xcolor}
\usepackage{fvextra}
\usepackage{float}
\newfloat{listing}{tbp}{lol}
\floatname{listing}{Listing}
\DefineVerbatimEnvironment{MintedVerbatim}{Verbatim}{}
\input{highlighted/tango-style.tex}

\usepackage{graphicx}

\newunicodechar{ℕ}{\ensuremath{\mathbb{N}}}
\newunicodechar{→}{\ensuremath{\to}}
\newunicodechar{≤}{\ensuremath{\leq}}
\newunicodechar{⟨}{\ensuremath{\langle}}
\newunicodechar{⟩}{\ensuremath{\rangle}}
\newunicodechar{↦}{\ensuremath{\mapsto}}
\newunicodechar{₁}{\ensuremath{{}_1}}
\newunicodechar{·}{\ensuremath{\cdot}}
\newunicodechar{⁻}{\ensuremath{{}^{-1}}}
\newunicodechar{∀}{\ensuremath{\forall}}
\newunicodechar{α}{\ensuremath{\alpha}}
\newunicodechar{₂}{\ensuremath{{}_2}}

\newcommand\solver[1]{\textsc{#1}\xspace}

\newcommand\vampire{\solver{Vampire}}
\newcommand\lean{\solver{Lean}}
\newcommand\avatar{\solver{Avatar}}
\newcommand\leanauto{\solver{Lean-auto}}
\newcommand\vampirehammer{\solver{Vampire-Hammer}}
\newcommand\aesop{\solver{Aesop}}
\newcommand\duper{\solver{Duper}}

\renewcommand{\paragraph}[1]{\par\smallskip\noindent\textbf{#1}}

\def\orcidID#1{\href{http://orcid.org/#1}{\raisebox{-1.25pt}{\includegraphics{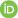}}}}
\begin{document}
\title{Hammering with Vampire in Lean}
\author{Jonas Bodingbauer\orcidID{0009-0006-2607-4325} \and
Laura Kov\'acs\orcidID{0000-0002-8299-2714}}
\authorrunning{Bodingbauer and Kov\'acs}
\institute{TU Wien, Vienna, Austria\\
\email{jonas.bodingbauer@tuwien.ac.at}\\
}
\maketitle              %
\begin{abstract}
We present \vampirehammer, a hammer for the \lean interactive theorem prover that discharges goals using the \vampire automated theorem prover and reconstructs proofs checked by the \lean kernel. \vampirehammer uses the %
\leanauto framework to process goals and premises before dispatching them to \vampire for proof search. 
Doing so, we rely on standard TPTP problem formats and annotated TSTP proofs that support proof reconstruction. With a \(10\)-second budget on 196,083 eligible Mathlib theorems, \vampirehammer solves 21,158 problems not solved by  other evaluated tools within the same budget.
\end{abstract}

\section{Introduction}
The \vampire theorem prover fully automates proof search in first- and higher-order logic, including theories and induction~\cite{DBLP:conf/cav/BartekBCHHHKRRRSSV25}. Its proving capabilities  support interactive theorem provers (ITPs) for higher-order logics: \vampire has long served as a proof hammer in Isabelle~\cite{DBLP:conf/tphol/WenzelPN08,sledgehammer}. However, integrating automated theorem provers (ATPs), particularly \vampire, as proof hammers with full reconstruction in ITPs based on dependent type theory, such as \lean~\cite{DBLP:conf/cade/Moura021}, remains challenging. Recent approaches translate fragments of dependent type theory into first- and higher-order logic~\cite{DBLP:conf/cav/MohamedMKBRQTB25,DBLP:conf/cav/QianCBA25}, enabling ATPs and SMT solvers to reason about them. However, translating a goal and reconstructing an external proof are distinct challenges and not yet fully addressed by existing techniques. {This tool demonstration paper addresses these challenges with the new \vampirehammer tool}, which uses \vampire as a reasoning engine for \lean and fully reconstructs its proofs for kernel checking. Successful invocations of \vampirehammer thus do not require the translation procedure, the reconstruction or \vampire to be trusted for proof correctness.  To the best of our knowledge, \vampirehammer provides the first complete reasoning pipeline for (native) proof reconstruction,  supporting proof extraction in both first- and higher-order logic (Figure~\ref{fig:pipeline}). Furthermore, \vampirehammer provides the first hammer for \lean in a way that the proof-search efficiency of the ATP is fully maintained and reconstruction of proof-steps is shared later between both programs in an efficient way.

\begin{figure}[t]
\begin{minipage}[c]{0.63\linewidth}
\textbf{(a) Explicit proof}\par\smallskip
\begingroup\fvset{frame=single,framesep=1mm,fontsize=\scriptsize}\input{highlighted/listing-1.tex}\endgroup
\end{minipage}\hspace{1em}
\begin{minipage}[c]{0.3\linewidth}
\textbf{(b) Automatic tactic}\par\smallskip
\begingroup\fvset{frame=single,framesep=1mm,fontsize=\scriptsize}\input{highlighted/listing-2.tex}\endgroup
\end{minipage}
\caption{(a) A \lean goal expressing that {\tt centralSquares} imply central commutators, discharged by a single \vampirehammer call.
Here $G$ is a group and \texttt{Commute x y}
means $xy=yx$. The explicit proof factors the
commutator into three squares. (b) Invoked in \lean,  \vampirehammer
finds a proof from the {\tt centralSquares} hypothesis.}
\label{fig:theorem_example}
\end{figure}

\paragraph{Methodology.} Our work builds on prior results in proof checking~\cite{DBLP:conf/cade/RawsonVSK25,bodingbauer_et_al:LIPIcs.ITP.2026.36}, producing \vampire proofs in formats that can be checked by the SMT solver Z3~\cite{Z3} or by \lean. However, direct proof output alone is insufficient for a \vampire-based hammer in \lean: proof goals must (i) first be translated from dependent type theory into simply typed higher-order logic or, where possible, into typed first-order logic to be suitable for \vampire, and  (ii) the  resulting proof must then be reconstructed using the symbols and assumptions introduced during the translation.
Our \vampirehammer therefore combines the \leanauto translation framework~\cite{DBLP:conf/cav/QianCBA25} with proof reconstruction in \lean, and extends \vampire with annotated TSTP proof output, providing sufficient information for efficient reconstruction\footnote{also useful for other proof-processing tools}. Together, these components provide a complete pipeline in \vampirehammer (Section~\ref{sec:pipeline}): given an input \lean proof goal, we  reconstruct proof details in \vampire and handle all further proof-term generation in \lean which can then be checked by its kernel. Our work uses the TPTP syntax~\cite{TPTP} for problem exchange and the related TSTP format for proof output. Figure~\ref{fig:theorem_example} illustrates the use of \vampirehammer as a tactic (\texttt{vampire}) in \lean on a group theory theorem, showing how a nontrivial proof can be replaced by a single call to \vampirehammer.

We evaluated \vampirehammer on 196,083 Mathlib~\cite{Mathlib} theorems (Section~\ref{sec:evaluation}). Within a timeout of $10$ seconds, our experiments show that \vampirehammer solves 21,158 problems that no other tool in the benchmark solves within the same budget. 
Thanks to supported TPTP/TSTP standards, \vampirehammer could further be extended to support other ATPs, provided that their inference rules admit similar  simple reconstruction with extra annotations. Premise-selection tools, such as lean-hammer~\cite{zhu2025premiseselectionleanhammer}, complement our work and offer possible directions for future work (Section~\ref{sec:conclusion}).

\section{The \vampirehammer Pipeline}
\label{sec:pipeline}

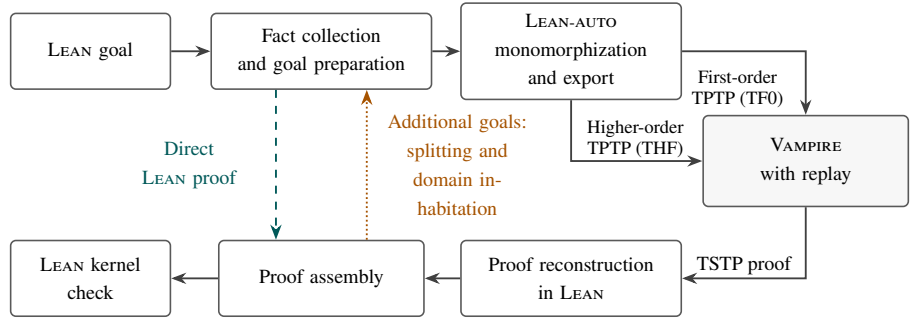
\begin{figure}[t]
    \centering
\begin{tikzpicture}[
  font=\fontsize{7.5}{10.8}\selectfont,
  flowbox/.style={draw=black!65, rounded corners=2pt,
    line width=.6pt, fill=white, align=center,
    minimum height=10mm, text width=27mm,
    inner xsep=3pt, inner ysep=4pt},
  flow/.style={-{Stealth[length=2mm]},
    line width=.7pt, draw=black!75},
  direct/.style={flow, dashed, draw=teal!70!black},
  additional/.style={flow, densely dotted,
    draw=orange!65!black}
]
  \node[flowbox, text width=19mm] (goal)
    at (0,3.0) {\lean goal};
  \node[flowbox] (prepare) at (3.05,3.0)
    {Fact collection\\and goal preparation};
  \node[flowbox] (auto) at (6.35,3.0)
    {\leanauto\\monomorphization\\and export};

  \node[flowbox, text width=25mm,
    minimum height=12mm, fill=black!3]
    (vampire) at (9.45,1.55)
    {\vampire\\{\fontsize{7.5}{9}\selectfont with replay}};

  \node[flowbox] (reconstruct) at (6.35,0)
    {Proof reconstruction\\in \lean};
  \node[flowbox, text width=25mm] (assemble)
    at (3.05,0) {Proof assembly};
  \node[flowbox, text width=19mm] (checked)
    at (0,0) {\lean kernel\\check};

  \draw[flow] (goal) -- (prepare);
  \draw[flow] (prepare) -- (auto);

  \draw[flow] (auto.south) --
    (6.35,1.55) -- (vampire.west);
  \node[anchor=south, align=center, inner sep=1pt,
    font=\fontsize{7}{8}\selectfont]
    at (7.2,1.56) {Higher-order\\TPTP (THF)};

  \draw[flow] (auto.east) --
    (9.45,3.0) -- (vampire.north);
  \node[anchor=south, align=center, inner sep=1pt,
    font=\fontsize{7}{8}\selectfont]
    at (8.55,2.22) {First-order\\TPTP (TF0)};

  \draw[flow] (vampire.south) -- (9.45,0) --
    node[midway, above, inner sep=1pt]
      {TSTP proof}
    (reconstruct.east);

  \draw[flow] (reconstruct) -- (assemble);
  \draw[flow] (assemble) -- (checked);

  \draw[direct] ([xshift=-6mm]prepare.south) --
    node[midway, left=1.5mm,
      text width=19mm, align=center, inner sep=1pt,
      fill=white, text=teal!70!black]
      {Direct \lean proof}
    ([xshift=-6mm]assemble.north);

  \draw[additional]
    ([xshift=6mm]assemble.north) --
    node[midway, right=1.5mm,
      text width=20mm, align=center, inner sep=1pt,
      fill=white, text=orange!65!black]
      {Additional goals:\\splitting and\\domain inhabitation}
    ([xshift=6mm]prepare.south);
\end{tikzpicture}
\caption{The \vampirehammer pipeline via \leanauto and proof reconstruction. Additional goals from goal splitting and domain inhabitation return to goal preparation.\vspace*{-1em}}
\label{fig:pipeline}
\end{figure}

\vampirehammer yields a proof tactic that integrates the \vampire automated theorem prover
with \lean through a proof-producing translation and reconstruction
pipeline, as summarized in Figure~\ref{fig:pipeline} and detailed next. Given a \lean
goal and a collection of premises, the \vampirehammer tactic first attempts to
discharge the goal using lightweight \lean-native reasoning \footnote{reflexivity, applying collected facts, equality substitution, and constructor injectivity or disjointness}. If this
fails, \vampirehammer translates the input problem into a representation supported by
\vampire, invokes external proof search, and reconstructs the
resulting refutation as a \lean proof. All successful tactic
invocations ultimately produce proof terms checked by \lean's kernel.

\paragraph{Goal preparation and direct reasoning.}
Given a goal $G$ and a collection of selected premises
$\Gamma$, \vampirehammer reduces the task of proving $G$ to refuting $G$, by
introducing $\neg G$ and attempting to derive $\bot$ from
$\Gamma \cup \{\neg G\}$. The frontend collects explicit premises and automatically available facts,
including typeclass laws and defining equations, and normalizes the corresponding
\lean expressions through reductions and definition unfolding as needed
for translation.

\paragraph{Monomorphization and preprocessing.}
After goal preparation, the \vampirehammer frontend delegates monomorphization and translation to \leanauto~\cite{DBLP:conf/cav/QianCBA25}. Preprocessing also tracks inhabitation obligations: TPTP sorts are nonempty whereas \lean allows empty types. Depending on the configuration, preprocessing produces monomorphic higher- or first-order problem encodings.

\paragraph{External proof search.}
The resulting problem is then passed to \vampire either using  first-order logic (TPTP TF0) or higher-order logic (TPTP THF) encodings.  Source types, symbols, and formulas receive
fresh names, while the exporter retains mappings from these
names to their reified counterparts.

\vampire is invoked on the resulting problem, using portfolio mode for first-order logic and default mode for higher-order logic. When a proof is found, important inference information, such as unifiers and modified literal positions, are recovered using the replay mechanism introduced
in~\cite{bodingbauer_et_al:LIPIcs.ITP.2026.36}.
Listing~\ref{lst:tstp-annotations} shows these annotations for a subsumption resolution step of \vampire's proof calculus, that could occur during proving Figure~\ref{fig:theorem_example}. Standard \vampire treats clauses as
unordered sets and it changes the order of literals in clauses without recording that. To make such steps reconstructible in \vampirehammer, we extend \vampire to output
associativity and commutativity bridges recording the applied 
permutations.

\paragraph{Proof decoding and local reconstruction.}
The  TSTP proof generated by \vampire is parsed in \lean~\cite{Polaczek2026} and converted into an internal representation with decoded annotations for reconstruction. 
\begin{listing}[t]
\footnotesize
\begin{lstlisting}
tff(f3,plain, ![X : group] : (mul(X,inv(X)) = e | mul(inv(X),X) = e),
  inf(fw_res, [uni([subs(f2,[b(Y,X)])]), rem_lit(lit(f1,0))],[f1,f2])).
tff(f4,plain, ![X : group] : (mul(inv(X),X) = e | mul(X,inv(X)) = e),
  inf(ac,[perm([1,0])],[f3])).
\end{lstlisting}
\caption{A shortened example of a TSTP proof step %
with information for fast
reconstruction. A \texttt{uni} annotation records the substitution used in
the inference, and \texttt{rem\_lit} records the location of the removed literal in the parent clause.
The \texttt{ac} bridge in the second step records the applied permutation.\vspace*{-1.5em}}
\label{lst:tstp-annotations}
\end{listing}
Following the approach of~\cite{bodingbauer_et_al:LIPIcs.ITP.2026.36}, each proof step is replayed as a local lemma with premises
$P_1,\ldots,P_n$ and conclusion $C$. The backend constructs
a \lean term of the form \(
    P_1 \to \cdots \to P_n \to C, \)
using reconstruction procedures for each inference rule. Many steps can be proven by properly instantiating theorems, such as the resolution theorem in Listing~\ref{lst:Resolution_Theorem}, using the annotations emitted by \vampire. 

When a rule-specific reconstruction procedure does not apply, for example because of missing annotations or an unexpected inference rule, \texttt{grind}~\cite{10.1007/978-3-032-32589-1_7} is used as a fallback.
Symbols introduced during external reasoning, such as Skolem constants, \avatar split predicates, and named definitions, require additional
justification in the form of local \texttt{let} statements as demonstrated in previous work~\cite{bodingbauer_et_al:LIPIcs.ITP.2026.36}.
Once each step of the refutation has been reconstructed locally, the individual lemmas are combined into one full theorem yielding the empty clause \(
\bot\), and thus concluding the refutation. The resulting proof is passed  to \leanauto for assembly into a term that can be checked against the initial goal by the \lean kernel.

\begin{listing}[b]
    \centering
    \begingroup\fvset{framesep=1mm,fontsize=\scriptsize}\input{highlighted/listing-3.tex}\endgroup
    \caption{\lean theorem for binary resolution for resolving complementary literals and deriving the disjunction of remaining literals with clauses treated as Lists of Props.%
    }
    \label{lst:Resolution_Theorem}
\end{listing}

\section{Search Strategy}
\label{sec:search}
\paragraph{Split higher-order and first-order routes.}
As shown in Figure~\ref{fig:pipeline}, problem export in \vampirehammer is possible in higher- and first-order logic. The higher-order logic export keeps function types and preserves a more expressive subset of dependent type theory, whereas first-order proof search is more mature and better optimized in \vampire. Although first-order search may be faster, translating a \lean goal into first-order logic can produce larger problems and discard information retained by higher-order encodings, potentially making proofs harder to find. Thus, both approaches are viable.

\paragraph{Domain inhabitation.} 
While \leanauto already collects domain inhabitation evidence,
\vampirehammer performs an additional, time-limited inhabitation search.
When no inhabitant is found for a type, \vampire is nevertheless invoked
under the assumption that the type is inhabited, leaving its justification
as an open proof obligation. If \vampire finds a proof that does not use
this assumption, the proof can be reconstructed immediately. Otherwise,
\vampirehammer considers the complementary case in which the type is
empty and attempts to discharge the resulting goal. If successful, the
proofs of both cases are combined to establish the original goal.

\paragraph{Scheduling higher-order and first-order search.}
Higher- and first-order search are complementary, with each finding proofs that the other cannot. When running on a single thread, the two approaches are allocated time slices determined empirically using a small subset of benchmark problems. During the first \(20\%\) of \vampirehammer's search budget, \vampire runs in higher-order mode, as it tends to terminate quickly on unsuccessful attempts and typically finds successful proofs early. If higher-order search fails, the remaining budget is allocated to first-order search using \vampire's portfolio mode. In case the first-order proof is unsuccessful, a higher-order attempt with additional hints is started.
Goals that can be easily split, such as propositions of the form \(A \land B\), are identified before proof search. For goals of this form, a fraction of the total time budget is reserved ahead of time for attempting proof search on  resulting subgoals \footnote{In parallel mode, recommended for interactive sessions and on per default, these searches can run concurrently. }%

\section{Experimental Evaluation}
\label{sec:evaluation}

\paragraph{Benchmark.}
Our benchmark consists of the user-declared theorems of Mathlib, evaluated with the same \lean version v4.34.1.
For each theorem, we extract the direct theorem dependencies appearing
in its original proof and provide these as explicit premises to the
competing tactics, together with the theorem's local hypotheses.
Thus, our experiments primarily evaluate automated reasoning given
proof-derived premises, rather than end-to-end premise selection tackled
by other approaches~\cite{zhu2025premiseselectionleanhammer}. 
Our comparison uses the intersection of eligible cases
across all tools, totaling in 196,083 theorems.

To prevent access to the target theorem or declarations introduced
later in its source file, the harness reconstructs the environment
immediately before the command that originally declared the target.
Targets whose statements or original proofs reference constants
introduced by the same command are excluded.

\paragraph{Execution protocol.}
Our experiments run on compute nodes with two AMD EPYC 9335 CPUs at 3\,GHz and 1.5\,TB of memory. The nodes are managed by Slurm, and each node runs 64 \lean instances in parallel. To minimize \lean loading times and source replay, each worker processes batches of up to 16 theorems from the same defining module for the same tactic. We ran the full benchmark twice and report the arithmetic mean across both runs for the runtimes shown in the plots and the number of successfully proved theorems.\footnote{For each tool, the number of proved theorems differed by at most two between runs.}
Every tactic invocation receives a wall-clock proof budget of \(10\,\mathrm{s}\),
including premise extraction, tactic execution, and kernel checking.
We disable \lean's overall heartbeat limit for all tools.
This avoids terminating an otherwise admissible attempt solely
because of \lean's default heartbeat allowance. 
A success requires a complete proof accepted by \lean's kernel.
The harness also rejects proofs containing \texttt{sorryAx} or
depending on the target theorem itself.

\begin{figure}[t]
  \centering
  \includegraphics{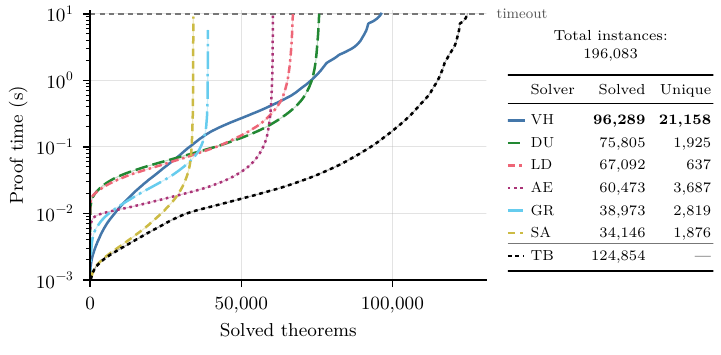}
  \caption{Cactus plots and solve counts over two runs with a \(10\,\mathrm{s}\) proof budget. Unique counts denote problems solved by the respective tool, not solved by any other tool in the suite. VH: \vampirehammer; LD: \leanauto + \duper; DU: \duper; AE: \aesop; SA: \texttt{simp\_all}; GR: \texttt{grind}; TB: theoretical best, combining all tools using the fastest successful proof per problem.\vspace*{-1em}}
  \label{fig:cactus-plot}
\end{figure}

\paragraph{Results.}
We compare \vampirehammer against \leanauto+\duper \cite{DBLP:conf/cav/QianCBA25,clune_et_al:LIPIcs.ITP.2024.10},
standalone \duper~\cite{clune_et_al:LIPIcs.ITP.2024.10}, \aesop~\cite{aesop}, \texttt{simp\_all}, and \texttt{grind}~\cite{10.1007/978-3-032-32589-1_7} \footnote{The respective benchmarked version commits are available in the appendix.}.
Out of 196,083 theorems, 96,289 are proven within \(10\,\mathrm{s}
\) by \vampirehammer.
As shown in Figure~\ref{fig:cactus-plot}, the union of all tools solve 124,854 theorems, with 21,158 unique problems 
solved only by \vampirehammer, showcasing the advantage of our proof automation and a relative improvement of \(20.4\%\). This demonstrates that \vampirehammer extends the reach of existing \lean proof automation, yielding an improvement to the existing automation landscape. Of all problems exported to \vampire, 62\% were solved using higher-order reasoning, 36\% using first-order reasoning, and 1\% using both. These numbers are largely influenced by the higher-order search occurring first in the employed schedule.
The noticeable kinks in the \vampirehammer curve  of Figure~\ref{fig:cactus-plot} reflect transitions between the proof search modes detailed in Section~\ref{sec:search}. 

\section{Conclusion}
\label{sec:conclusion}

We presented \vampirehammer, a proof-producing hammer that brings the
\vampire automated theorem prover to \lean. Given a goal and selected
premises, \vampirehammer exports  problems through \leanauto into
first-order or higher-order TPTP, searches for a refutation with
\vampire, and reconstructs the annotated TSTP proof inside \lean as a
composition of local lemmas, yielding a proof term checked by \lean's
kernel. Because proofs are exchanged through TSTP, the reconstruction approach
may be extended to other ATPs if their inference rules and proof output
can be supported. On 196,083 eligible
Mathlib~\cite{Mathlib} theorems, \vampirehammer solves 21,158
problems that no other evaluated tool solves within the same budget.
Future work is the integration of premise selection over the full
library, e.g.\ via lean-hammer~\cite{zhu2025premiseselectionleanhammer}, improving exported problem quality, and higher-order reasoning in \vampire. 

\paragraph{Acknowledgments.} We thank Henrik Böving for discussions and feedback on our work. %

\paragraph{Funding.}
This research was funded in whole or in part by the  Austrian Science Fund -- FWF [Grant ID 10.55776/EFP2661725], the WWTF grant ForSmart
10.47379/ICT22007, and the SBA Research  COMET Center SBA-K1 NGC managed  by the FFG.

\paragraph{AI Statement.} 
{Generative AI tools (GPT-6.1-Sol, Claude Opus 5.5 and GLM 5.3) were used to assist with the implementation and refinement of parts of the \vampirehammer.} %

\paragraph{Data Availability Statement.}
The implementation  for \vampirehammer is available at \url{https://github.com/vprover/vampire-hammer}, with accompanying benchmark data in \url{https://doi.org/10.5281/zenodo.23405497} as summarized in Appendix~\ref{app:B}.

\bibliographystyle{splncs04}
\bibliography{refs}

\newpage
\appendix
\section{Benchmark Setup}
\label{app:benchmark-setup}
The benchmark evaluates each theorem in an environment reconstructed from immediately before its declaration, excluding the target itself and later declarations. Although every tool receives the same explicit premise names extracted from the original proof, these premises do not form an exclusive set of premises: tactics may also use global simplification rules, registered \texttt{grind} rules, and typeclass instances available in that environment. Such background knowledge can make some goals simpler, including cases where an earlier theorem already provides an equivalent result. The results therefore measure automation with declaration-time library support, rather than reasoning solely from the extracted premises.
\paragraph{Scope and limitations.}
The evaluation measures reconstruction of existing Mathlib theorems
with premises derived from their original proofs.
It does not measure interactive usability or premise discovery
from the full library.

\section{Benchmarked Tool Versions}\label{app:B}
\begin{table}[htbp]
\centering
\small
\begin{tabular}{lll}
\hline
Tool or component & Version & Commit \\
\hline
\vampirehammer & 0.1.0 & \texttt{4c50a21de3e4} \\
\vampire & 5.1.0 &       \texttt{6ee1f120df1b} \\
\duper & 4.34.1 & \texttt{9e09cd05fd54} \\
\leanauto  & 4.34.1 & \texttt{6b2774cde4d1} \\
\aesop & 4.34.0 & \texttt{355695d523e4} \\
\texttt{simp\_all} (Lean) & 4.34.1 & \texttt{5045d0056413} \\
\texttt{grind} (Lean) & 4.34.1 & \texttt{5045d0056413} \\
\hline
Mathlib & 4.34.1 & \texttt{d13f23b723b8} \\
\leanauto fork used by Vampire-Hammer & --- & \texttt{a2f825aca7e5} \\
\hline
\end{tabular}
\caption{Tool versions and dependency pins from the saved benchmark
setup, with abbreviated commit identifiers. A dash indicates that no release version
was established.}
\label{tab:tool-versions}
\end{table}
The proof attempt results of each tool in each run are available as two CSV files which each contain 196,083 theorems eligible.
The benchmarking scripts are available under \url{https://github.com/i-am-a-teapot/vampire-hammer-benchmarks}

\section{Usage}

\vampirehammer is installed as a Lake dependency, which automatically
builds Vampire and the bundled TPTP parser. It can be installed by a simple addition to the \lean project's lakefile like:
\begingroup\fvset{frame=single,framesep=1mm,fontsize=\footnotesize}\input{highlighted/listing-4.tex}\endgroup

\noindent More detailed information is available in the readme of the Git repository.

After importing
\texttt{VampireHammer}, users can invoke the \texttt{vampire} tactic on
selected hypotheses or lemmas like any other \lean tactic.
\begingroup\fvset{frame=single,framesep=1mm,fontsize=\footnotesize}\input{highlighted/listing-5.tex}\endgroup
Alternatively, \texttt{vampire [*]} selects all local hypotheses. The central squares example is provided as a further example in the repository. 

By default, first-order and native higher-order search run concurrently,
with a search allowance of 10 seconds per attempt. Users can adjust
this allowance with \texttt{(timeout := 20)}, select sequential search
with \texttt{(parallel := false)}, and guide definition handling through
\texttt{(unfold := [f])} and \texttt{(defeq := [f])}.
Completed proofs are cached within the Lean session; setting
\texttt{(cache := false)} forces a fresh search and \texttt{vampire.outputPath} retains input and prover output.

\end{document}

%% file: highlighted/listing-1.tex
\begin{MintedVerbatim}[commandchars=\\\{\}]
\PYG{k+kn}{theorem}\PYG{+w}{ }\PYG{n}{centralSquares}\PYG{+w}{ }\PYG{o}{(}\PYG{n}{h}\PYG{+w}{ }\PYG{o}{:}\PYG{+w}{ }\PYG{n+nb+bp}{∀}\PYG{+w}{ }\PYG{n}{x}\PYG{+w}{ }\PYG{n}{y}\PYG{+w}{ }\PYG{o}{:}\PYG{+w}{ }\PYG{n}{G}\PYG{o}{,}\PYG{+w}{ }\PYG{n}{Commute}\PYG{+w}{ }\PYG{o}{(}\PYG{n}{x}\PYG{n+nb+bp}{*}\PYG{n}{x}\PYG{o}{)}\PYG{+w}{ }\PYG{n}{y}\PYG{o}{)}
\PYG{+w}{    }\PYG{o}{(}\PYG{n}{a}\PYG{+w}{ }\PYG{n}{b}\PYG{+w}{ }\PYG{n}{c}\PYG{+w}{ }\PYG{o}{:}\PYG{+w}{ }\PYG{n}{G}\PYG{o}{)}\PYG{+w}{ }\PYG{o}{:}\PYG{+w}{ }\PYG{n}{Commute}\PYG{+w}{ }\PYG{o}{(}\PYG{n}{a}\PYG{+w}{ }\PYG{n+nb+bp}{*}\PYG{+w}{ }\PYG{n}{b}\PYG{+w}{ }\PYG{n+nb+bp}{*}\PYG{+w}{ }\PYG{n}{a}\PYG{n+nb+bp}{⁻}\PYG{+w}{ }\PYG{n+nb+bp}{*}\PYG{+w}{ }\PYG{n}{b}\PYG{n+nb+bp}{⁻}\PYG{o}{)}\PYG{+w}{ }\PYG{n}{c}\PYG{+w}{ }\PYG{o}{:=}\PYG{+w}{ }\PYG{k}{by}
\PYG{+w}{  }\PYG{k}{have}\PYG{+w}{ }\PYG{n}{factor}\PYG{+w}{ }\PYG{o}{:}\PYG{+w}{ }\PYG{n}{a}\PYG{+w}{ }\PYG{n+nb+bp}{*}\PYG{+w}{ }\PYG{n}{b}\PYG{+w}{ }\PYG{n+nb+bp}{*}\PYG{+w}{ }\PYG{n}{a}\PYG{n+nb+bp}{⁻}\PYG{+w}{ }\PYG{n+nb+bp}{*}\PYG{+w}{ }\PYG{n}{b}\PYG{n+nb+bp}{⁻}\PYG{+w}{ }\PYG{n+nb+bp}{=}
\PYG{+w}{    }\PYG{c+c1}{\PYGZhy{}\PYGZhy{}...}
\PYG{+w}{  }\PYG{n}{rw}\PYG{+w}{ }\PYG{o}{[}\PYG{n}{factor}\PYG{o}{]}
\PYG{+w}{  }\PYG{n}{exact}\PYG{+w}{ }\PYG{o}{((}\PYG{n}{h}\PYG{+w}{ }\PYG{n}{a}\PYG{+w}{ }\PYG{n}{c}\PYG{o}{)}\PYG{n+nb+bp}{.}\PYG{n}{mul\PYGZus{}left}
\PYG{+w}{    }\PYG{o}{(}\PYG{n}{h}\PYG{+w}{ }\PYG{o}{(}\PYG{n}{a}\PYG{n+nb+bp}{⁻}\PYG{+w}{ }\PYG{n+nb+bp}{*}\PYG{+w}{ }\PYG{n}{b}\PYG{o}{)}\PYG{+w}{ }\PYG{n}{c}\PYG{o}{))}\PYG{n+nb+bp}{.}\PYG{n}{mul\PYGZus{}left}\PYG{+w}{ }\PYG{o}{(}\PYG{n}{h}\PYG{+w}{ }\PYG{n}{b}\PYG{n+nb+bp}{⁻}\PYG{+w}{ }\PYG{n}{c}\PYG{o}{)}
\end{MintedVerbatim}

%% file: highlighted/listing-2.tex
\begin{MintedVerbatim}[commandchars=\\\{\}]
\PYG{k+kn}{import}\PYG{+w}{ }\PYG{n}{VampireHammer}

\PYG{k+kn}{theorem}\PYG{+w}{ }\PYG{n}{centralSquares}\PYG{+w}{ }\PYG{n+nb+bp}{...}
\PYG{+w}{    }\PYG{o}{:=}\PYG{+w}{ }\PYG{k}{by}\PYG{+w}{ }\PYG{n}{vampire}
\end{MintedVerbatim}

%% file: highlighted/listing-3.tex
\begin{MintedVerbatim}[commandchars=\\\{\}]
\PYG{k+kn}{theorem}\PYG{+w}{ }\PYG{n}{resolution}\PYG{+w}{ }\PYG{o}{(}\PYG{n}{lit}\PYG{+w}{ }\PYG{o}{:}\PYG{+w}{ }\PYG{k+kt}{Prop}\PYG{o}{)}\PYG{+w}{ }\PYG{o}{(}\PYG{n}{c1}\PYG{+w}{ }\PYG{n}{c1\PYGZsq{}}\PYG{+w}{ }\PYG{n}{c2}\PYG{+w}{ }\PYG{n}{c2\PYGZsq{}}\PYG{+w}{ }\PYG{o}{:}\PYG{+w}{ }\PYG{n}{List}\PYG{+w}{ }\PYG{k+kt}{Prop}\PYG{o}{)}\PYG{+w}{ }\PYG{o}{:}\PYG{+w}{ }\PYG{n}{H}\PYG{+w}{ }\PYG{o}{(}\PYG{n}{c1}\PYG{+w}{ }\PYG{n+nb+bp}{++}\PYG{+w}{ }\PYG{n}{lit}\PYG{+w}{ }\PYG{n+nb+bp}{::}\PYG{+w}{ }\PYG{n}{c1\PYGZsq{}}\PYG{o}{)}\PYG{+w}{ }\PYG{n+nb+bp}{→}
\PYG{+w}{    }\PYG{n}{H}\PYG{+w}{ }\PYG{o}{(}\PYG{n}{c2}\PYG{+w}{ }\PYG{n+nb+bp}{++}\PYG{+w}{ }\PYG{o}{(}\PYG{n+nb+bp}{¬}\PYG{+w}{ }\PYG{n}{lit}\PYG{o}{)}\PYG{+w}{ }\PYG{n+nb+bp}{::}\PYG{+w}{ }\PYG{n}{c2\PYGZsq{}}\PYG{o}{)}\PYG{+w}{ }\PYG{n+nb+bp}{→}\PYG{+w}{ }\PYG{n}{H}\PYG{+w}{ }\PYG{o}{((}\PYG{n}{c1}\PYG{+w}{ }\PYG{n+nb+bp}{++}\PYG{+w}{ }\PYG{n}{c1\PYGZsq{}}\PYG{o}{)}\PYG{+w}{ }\PYG{n+nb+bp}{++}\PYG{+w}{ }\PYG{o}{(}\PYG{n}{c2}\PYG{+w}{ }\PYG{n+nb+bp}{++}\PYG{+w}{ }\PYG{n}{c2\PYGZsq{}}\PYG{o}{))}
\end{MintedVerbatim}

%% file: highlighted/listing-4.tex
\begin{MintedVerbatim}[commandchars=\\\{\}]
\PYG{n}{require}\PYG{+w}{ }\PYG{n}{vampire\PYGZus{}hammer}\PYG{+w}{ }\PYG{k}{from}
\PYG{+w}{    }\PYG{n}{git}\PYG{+w}{ }\PYG{l+s+s2}{\PYGZdq{}https://github.com/vprover/vampire\PYGZhy{}hammer\PYGZdq{}}\PYG{+w}{ }\PYG{n+nb+bp}{@}\PYG{+w}{ }\PYG{l+s+s2}{\PYGZdq{}main\PYGZdq{}}
\end{MintedVerbatim}

%% file: highlighted/listing-5.tex
\begin{MintedVerbatim}[commandchars=\\\{\}]
\PYG{k+kd}{example}\PYG{+w}{ }\PYG{o}{\PYGZob{}}\PYG{n}{α}\PYG{+w}{ }\PYG{o}{:}\PYG{+w}{ }\PYG{k+kt}{Type}\PYG{o}{\PYGZcb{}}\PYG{+w}{ }\PYG{o}{(}\PYG{n}{p}\PYG{+w}{ }\PYG{n}{q}\PYG{+w}{ }\PYG{o}{:}\PYG{+w}{ }\PYG{n}{α}\PYG{+w}{ }\PYG{n+nb+bp}{→}\PYG{+w}{ }\PYG{k+kt}{Prop}\PYG{o}{)}\PYG{+w}{ }\PYG{o}{(}\PYG{n}{a}\PYG{+w}{ }\PYG{o}{:}\PYG{+w}{ }\PYG{n}{α}\PYG{o}{)}
\PYG{+w}{    }\PYG{o}{(}\PYG{n}{h₁}\PYG{+w}{ }\PYG{o}{:}\PYG{+w}{ }\PYG{n+nb+bp}{∀}\PYG{+w}{ }\PYG{n}{x}\PYG{o}{,}\PYG{+w}{ }\PYG{n}{p}\PYG{+w}{ }\PYG{n}{x}\PYG{+w}{ }\PYG{n+nb+bp}{→}\PYG{+w}{ }\PYG{n}{q}\PYG{+w}{ }\PYG{n}{x}\PYG{o}{)}\PYG{+w}{ }\PYG{o}{(}\PYG{n}{h₂}\PYG{+w}{ }\PYG{o}{:}\PYG{+w}{ }\PYG{n}{p}\PYG{+w}{ }\PYG{n}{a}\PYG{o}{)}\PYG{+w}{ }\PYG{o}{:}\PYG{+w}{ }\PYG{n}{q}\PYG{+w}{ }\PYG{n}{a}\PYG{+w}{ }\PYG{o}{:=}\PYG{+w}{ }\PYG{k+kd}{by}
\PYG{+w}{  }\PYG{n}{vampire}\PYG{+w}{ }\PYG{o}{[}\PYG{n}{h₁}\PYG{o}{,}\PYG{+w}{ }\PYG{n}{h₂}\PYG{o}{]}
\end{MintedVerbatim}